\documentclass[pra,aps,twocolumn]{revtex4}
\usepackage{bm}
\usepackage{graphicx}
\usepackage{amsbsy}
\usepackage{amsmath}
\usepackage{amsfonts}

\newcommand{\bra}[1]{\langle{#1}|}
\newcommand{\ket}[1]{|{#1}\rangle}
\newcommand{\braket}[2]{\langle{#1}|{#2}\rangle}
\newcommand{\nn}{\nonumber}
\newcommand{\dg}{^\dagger}
\newcommand{\Alpha}{{\rm A}}

\begin{document}

\title{Resources required for exact remote state preparation}

\author{Dominic W.\ Berry$^{1,2}$}
\affiliation{$^1$ Australian Centre for Quantum Computer Technology,
Department of Physics, Macquarie University, Sydney, New South Wales
2109, Australia \\
$^2$ Institute for Quantum Information Science, Department of Physics and
Astronomy, University of Calgary, Alberta T2N 1N4, Canada}

\begin{abstract}
It has been shown [M.-Y.\ Ye, Y.-S.\ Zhang, and G.-C.\ Guo, \pra {\bf 69},
022310 (2004)] that it is possible to perform exactly faithful remote state
preparation using finite classical communication and any entangled state with
maximal Schmidt number. Here we give an explicit procedure for performing this
remote state preparation. We show that the classical communication required for
this scheme is close to optimal for remote state preparation schemes of this
type. In addition we prove that it is necessary that the resource state have
maximal Schmidt number.
\end{abstract}
\date{\today}

\maketitle

\section{Introduction}
Remote state preparation (RSP) is the preparation of a state at a remote
location using entanglement and classical communication
\cite{lo,pati,bennett,devetak,berry,ye,benn2}. In general, one may
perform exactly faithful RSP \cite{pati,bennett,ye}, producing exactly the
desired state, or asymptotically faithful RSP, where the fidelity approaches
one as the number of states prepared approaches infinity
\cite{lo,bennett,devetak,berry,benn2}.

It is well known that it is not possible to perform exactly faithful RSP without
entanglement. An infinite amount of classical information is required to exactly
represent an arbitrary state, and therefore exact RSP would require an infinite
amount of classical communication if there were no entangled resource. A method
for exact RSP of a restricted ensemble of states is given in Ref.\ \cite{pati},
and an alternative method for exact RSP of arbitrary states is given in
Ref.\ \cite{bennett}. Recently Ye {\it et al.}\ \cite{ye} showed that it is
possible to perform exact RSP using any pure entangled state, provided the
Schmidt number is equal to the system dimension. However, the proof given in
Ref.\ \cite{ye} does not give a complete technique for performing this remote
state preparation.

Here we give an explicit technique that is based upon an approximate technique
without entanglement, and quantify how much classical communication is required
for this scheme. In Sec.\ \ref{sec:scheme} we describe this scheme, then we
discuss the classical communication required in Sec.\ \ref{sec:comm}. We show
that the initial entangled state must have maximal Schmidt number in
Sec.\ \ref{sec:schmidt}, and conclude in Sec.\ \ref{sec:conc}.

\section{Explicit Scheme}
\label{sec:scheme}
As in Ref.\ \cite{ye}, the initial state is an entangled state of the form
\begin{equation}
\ket{\Alpha} = \sum_{k=0}^{d-1} \alpha_k \ket{k}\ket{k},
\end{equation}
where the $\alpha_k$ are nonzero real numbers, and each subsystem is of
dimension $d$. Any entangled state with maximal Schmidt number may be brought
to this form via local operations. The state we wish to prepare is
\begin{equation}
\ket{\beta} = \sum_{k=0}^{d-1} \beta_k \ket{k},
\end{equation}
where the $\beta_k$ may be complex.

To explain this remote state preparation scheme, we first explain a simple
approximate scheme that one would use if no entangled resource state were
available. In this case, one would communicate an approximation of the
coefficients $\beta_k$, and prepare a state based on those coefficients. To
approximate $\beta_k$, note that the real and imaginary parts of $\beta_k$ will
be numbers in the interval $[-1,1]$. We can approximate $\beta_k$ by dividing
the interval $[-1,1]$ into $D$ subintervals
\begin{equation}
[-1,2/D-1), ~ [2/D-1,4/D-1), ~ \cdots ~ , [1-2/D,1].
\end{equation}
We then denote the numbers of the subintervals that the real and imaginary
parts of $\beta_k$ lie in as $n^r_k$ and $n^c_k$. That is,
\begin{align}
n^r_k &= \min\{D,\lfloor D (\text{Re} \beta_k+1)/2 \rfloor+1\}, \nn \\
n^c_k &= \min\{D,\lfloor D (\text{Im} \beta_k+1)/2 \rfloor+1\}.
\end{align}
The $\min$ takes account of the fact that the last subinterval is closed, so 1
lies in subinterval $D$. We may then approximate the real and imaginary parts
of $\beta_k$ as
\begin{equation}
\text{Re}\beta_k \approx (2n^r_k-1)/D-1, \quad \text{Im}\beta_k \approx
(2n^c_k-1)/D-1.
\end{equation}
The error in this approximation will be no more than $1/D$.

We may define a state corresponding to this approximation by
\begin{equation}
\ket{\tilde\beta'} = \sum_{k=0}^{d-1} \{(2n^r_k-1)/D-1+i[(2n^c_k-1)/D-1]\}
\ket{k}.
\end{equation}
This state will satisfy
\begin{equation}
\label{distance}
\big\| \ket{\beta}-\ket{\tilde\beta'}\big\|\le \frac{\sqrt{2d}}D.
\end{equation}
However, the state $\ket{\tilde\beta'}$ is not necessarily normalized; the state
that is prepared will be the corresponding normalized state, $\ket{\beta'}$.
This state may be a slightly poorer approximation, but will still satisfy (see
Appendix \ref{dist})
\begin{equation}
\label{fidelity}
\big| \braket{\beta}{\beta'}\big|^2\ge 1-\frac{2d}{D^2}.
\end{equation}

Without an entangled state, one would communicate the $2d$ numbers $n^r_k$ and
$n^c_k$ using $2d\log D$ bits. Here we use the convention that $\log$ indicates
logarithms base 2. We also use the convention that the number of ``bits'' is
the logarithm base 2 of the number of messages, and need not be an integer. The
preparer would intialize the system in the state $\ket 0$, then apply a unitary
operation $U$ such that the final state is $\ket{\beta'}$.

In the case where an entangled state is available, one may initialize the
system in an alternative state $\ket{\psi}$ that is close to $\ket 0$, such
that the operation $U$ takes the system to the exact state $\ket{\beta}$. We
express the required initial state $\ket{\psi}$ as
\begin{equation}
\ket{\psi} = \sum_{k=0}^{d-1} \psi_k e^{i\varphi_k} \ket k.
\end{equation}
In order to prepare this state, we first apply an entanglement transformation
scheme to transform the entangled state $\ket{\Alpha}$ to a second state
\begin{equation}
\ket{\Psi} = \sum_{k=0}^{d-1} \psi_k \ket k \ket k.
\end{equation}
The communication that is required depends on the entanglement transformation
method that is used. There are a number of different methods of performing
entanglement transformations \cite{nielsen,nielvid,jensen}, but there is the
problem that most of these methods require local operations in subsystem 2 that
are dependent on the state to be prepared.

It is possible to use the entanglement transformation scheme in
Ref.\ \cite{jensen}, though this method requires communication of $\log d!$
bits to communicate the permutation used. Via Caratheodory's theorem one may
restrict the number of possible permutations to $d^2-2d+2$, indicating that the
communication required is approximately $2\log d$. However, the set of
$d^2-2d+2$ permutations is dependent on the state to be prepared, so it is
still necessary to communicate $\log d!$ bits.

Here we describe a straightforward method of determining a set of permutations
that is independent of the state to be prepared. In general, in order to
perform the entanglement transformation, it is necessary that
$\vec\alpha^2\prec\vec\psi^2$. Here we apply the slightly stronger condition
that $\psi_0^2$ is greater than $1-r^2(d-1)$, where $r=\min\{\alpha_i\}$.
This condition implies that the majorization relation holds (see Appendix
\ref{major}).

The entanglement transformation may be achieved via a two step process. First
the state is transformed from $\ket{\Alpha}$ to the intermediate state
\begin{equation}
\ket{\Phi} = \sum_{k=0}^{d-1} \phi_k \ket{k} \ket{k},
\end{equation}
where $\phi_0=\psi_0$ and $\phi_k=\sqrt{(1-\psi_0^2)/(d-1)}$ for $k>0$. This
entanglement transformation may be achieved using the measurement operators
\begin{equation}
\label{measop1}
A_k = \sqrt{p_k} \left( \sum_{l=1;l\ne k}^{d-1} \frac{\phi_l}{\alpha_l}
\ket{l}\bra{l} + \frac{\phi_k}{\alpha_0}\ket{0}\bra{0}
+ \frac{\phi_0}{\alpha_k}\ket{k}\bra{k}\right),
\end{equation}
for $k>0$, and
\begin{equation}
\label{measop1a}
A_0 = \sqrt{p_0} \left( \sum_{l=1}^{d-1} \frac{\phi_l}{\alpha_l} \ket{l}\bra{l}
+ \frac{\phi_0}{\alpha_0}\ket{0}\bra{0} \right).
\end{equation}
The probabilities $p_k=(\alpha_k^2-\phi_k^2)/(\phi_0^2-\phi_k^2)$ for $k>0$ and
$p_0=1-\sum_{k>0}p_k$. On obtaining measurement result $k$, if $k>0$ it is
necessary to swap states $\ket{0}$ and $\ket{k}$. The total number of
measurement results is $d$, so the communication required is $\log d$.

This entanglement transformation is followed by an entanglement transformation
to take the state from $\ket{\Phi}$ to $\ket{\Psi}$. In this case the
measurement operators required are
\begin{equation}
\label{measop2}
B_k = \frac 1{\sqrt{d-1}} \left( \ket{0}\bra{0}+\sum_{l=1}^{d-1}
\frac{\psi_{l\oplus k}}{\phi_l} \ket{l}\bra{l}\right),
\end{equation}
where $k>0$ (there is no measurement operator for $k=0$). The notation $\oplus$
is used to indicate addition modulo $d-1$ but excluding 0
(i.e.\ $1+[(l+k-1) \mod (d-1)]$).
On obtaining measurement result $k$, it is necessary to perform
a cyclic permutation of the states $\ket{1}$ to $\ket{d-1}$. The total number
of possible measurement results is $d-1$, so the communication required is
$\log(d-1)$. Thus this method allows one to transform $\ket{\Alpha}$ to
$\ket{\Psi}$ with communication of only $\log(d^2-d)$.

One may then use the method applied in the proof of Theorem 1 of
Ref.\ \cite{ye} to obtain the state $\ket{\psi}$. That is, one may apply the
projection operators
\begin{equation}
P_k = \frac 1d \ket{\chi_k}\bra{\chi_k}
\end{equation}
where
\begin{equation}
\ket{\chi_k}=\sum_l e^{i[(2\pi/d)kl-\varphi_l]}\ket{l}.
\end{equation}
Upon obtaining measurement result $k$ one performs the local operation
\begin{equation}
C_k=\sum_l e^{i(2\pi/d)kl}\ket{l}\bra{l}.
\end{equation}
This step requires an additional $\log d$ bits of classical communication.

The final step is to perform the local operation in subsystem 1 to take the
state from $\ket{\psi}$ to $\ket{\beta}$. Communication of the numbers $n_k^r$
and $n_k^c$ that specify this operation requires communication of $2d\log D$.
To determine the value of $D$ necessary, note that we have required
$\psi_0^2\ge 1-r^2(d-1)$ in order to perform the entanglement transformation.
As $\psi_0^2=|\braket{0}{\psi}|^2=|\braket{\beta'}{\beta}|^2$, $\psi_0^2$ is
equal to the fidelity between the state to be prepared, $\ket{\beta}$, and the
approximate state $\ket{\beta'}$. From Eq.\ \eqref{fidelity}, the condition
$\psi_0^2\ge 1-r^2(d-1)$ will be satisfied for
\begin{equation}
D = \left\lceil \sqrt {\frac {2d}{r^2(d-1)} } \right\rceil .
\end{equation}

To summarize, the RSP scheme with entanglement is a three step process: \\
{\bf Step 1}: Transform $\ket{\Alpha}$ to $\ket{\Psi}$ using the measurement
operators \eqref{measop1}, \eqref{measop1a} and \eqref{measop2}. The
communication required is $\log(d^2-d)$. \\
{\bf Step 2}: Apply the method given in the proof of Theorem 1 of
Ref.\ \cite{ye} to prepare the unentangled state $\ket{\psi}$. This step
requires $\log d$ bits of communication. \\
{\bf Step 3}: Perform the unitary operation $U$ to transform $\ket{\psi}$ to
$\ket{\beta}$. This step requires communication of the numbers $n^r_k$ and
$n^c_k$ to determine the operation $U$, and therefore requires communication of
$2d \log D$ bits. \\

\section{Classical Communication Required}
\label{sec:comm}
The total classical communication for this scheme is approximately $3\log d +
2d \log D$. The classical communication required for this scheme is least when
the entangled state used is close to a maximally entangled state. The amount of
classical communication required goes to infinity as the entanglement
approaches zero; there is therefore a tradeoff, just as in the asymptotic
schemes considered by Refs.\ \cite{bennett,devetak}.

\begin{figure}
\centering
\includegraphics[width=0.45\textwidth]{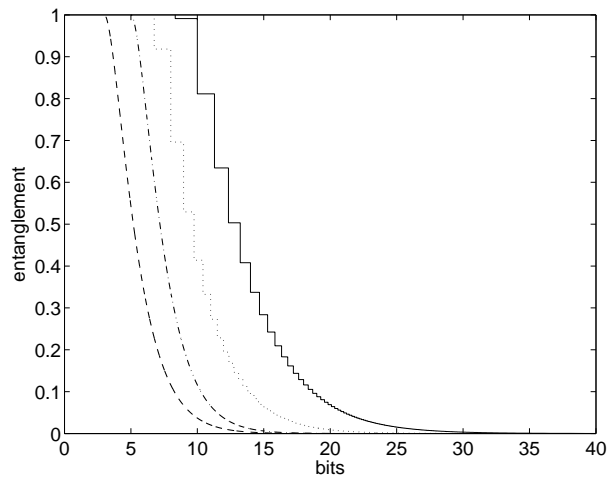}
\caption{The entanglement versus classical communication for exact RSP of qubit
states using a partially entangled state. The solid curve is that based on
the first scheme given, and the dotted line gives the communication required
when $\beta_0$ is taken to be real. The dash-dotted line is the upper bound on
the communication for the non-constructive scheme, and the dashed line
is a lower bound on the communication.}
\label{fig}
\end{figure}

The classical communication required is shown in Fig.\ \ref{fig} for the case
of a qubit. Comparing with the figure given in Refs.\ \cite{bennett,devetak},
we can see that the classical communication is significantly larger than for
asymptotically faithful RSP. In contrast to the asymptotic case, it is also
possible for the classical communication to approach infinity even if the
entanglement is not approaching zero. This is possible because one of the
Schmidt coefficients can become arbitrarily small even if the entanglement does
not.

One question that naturally arises is whether it is possible to perform this
RSP scheme with less classical communication. The total classical communication
required for steps 1 and 2 only scales as $\log d$. This communication is
already small, and it is unlikely that it can be improved upon. However, the
communication for the final step is $2d\log D$, which is much larger.

One may slightly reduce the communication required for step 3 by noting that
the global phase is arbitrary, so we may take $\beta_0$ to be real. Then it is
only necessary to approximate $2d-1$ numbers, and we obtain the fidelity
\begin{equation}
\big| \braket{\beta}{\beta'}\big|^2\ge 1-\frac{2d-1}{D^2}.
\end{equation}
Then the slightly lower value of $D$ may be taken
\begin{equation}
D = \left\lceil \sqrt {\frac {2d-1}{r^2(d-1)} } \right\rceil ,
\end{equation}
and the total communication for step 3 is $(2d-1)\log D$. This only gives a
slight reduction in the communication required; an example for qubit states is
given in Fig. \ref{fig}.

It is also possible to use a more efficient coding of the state. One
method is to record the sign of the real and imaginary parts of $\beta_k$, then
use $n_k^r$ and $n_k^c$ to approximate the absolute values of ${\rm Re}\beta_k$
and ${\rm Im}\beta_k$. For large $d$ most of the $n_k^r$ and $n_k^c$ will be
small, so it is more efficient to record the number of digits in the binary
representations of $n_k^r$ and $n_k^c$, as well as those digits. The total
communication required is then no more than (see Appendix \ref{efficient})
\begin{equation}
\label{second}
(2d-1)\left[ -\log (r\sqrt{d-1}) + \log \lceil \log D' \rceil + 2\right],
\end{equation}
where
\begin{equation}
D' = \left\lceil \sqrt {\frac {2d-1}{4r^2(d-1)} } \right\rceil .
\end{equation}
The first term is the communication required for the digits, and the second
term is the communication required for the numbers of digits. The third term
includes a correction for rounding, as well as the communication required for
the signs.

In assessing the scaling of each of the terms with $d$ it is necessary to assume
a scaling for $r$. It is not possible to take $r$ to be independent of $d$,
because $r\le 1/\sqrt d$. If $r\propto 1/\sqrt d$, the first term in
Eq.\ \eqref{second} scales approximately linearly with $d$, whereas the second
term scales as $d\log\log d$, and therefore is dominant for large $d$. However,
this situation is unlikely, because it would mean that the communication
required for the number of digits in $n_k^r$ and $n_k^c$ is less than that for
the digits themselves. It is more realistic to assume that $r$ decreases more
rapidly than $1/\sqrt d$ (for example as $1/d$); this is because, for larger
dimension, it is more likely that one of the Schmidt coefficients is
exceptionally small. Under this assumption, the first term is dominant, as would
be expected.

It is possible to perform the coding more efficiently than this, although the
proof is not constructive. In general, in order to approximate a state with
fidelity $1-\epsilon^2$, it is necessary to have a set of states
${\cal M}=\{\ket{\varphi_k}\}$, such that for any state $\ket{\beta}$, the
fidelity between $\ket{\beta}$ and some element of ${\cal M}$ is at least
$1-\epsilon^2$. To approximate the state, it is necessary to communicate the
index $k$ of a state that has fidelity at least $1-\epsilon^2$ with
$\ket{\beta}$. It was shown in Ref.\ \cite{benn2} that the number of states in
${\cal M}$ need be no greater than $(2.5/\epsilon)^{2d}$; here we apply a
similar method to improve upon this bound.

Consider a set ${\cal M}$ that satisfies the condition that
$|\braket{\varphi_k}{\varphi_l}|^2<1-\epsilon^2$ for $k\ne l$. The largest set
satisfying this condition, ${\cal M}_{\rm max}$, must also satisfy the fidelity
condition. This is because, if any state $\ket{\beta}$ satisfied
$|\braket{\varphi_k}{\beta}|^2<1-\epsilon^2$ for all $k$, it could be added
and thereby increase the size of the set.
Because no two states in ${\cal M}_{\rm max}$ have fidelity as large as
$1-\epsilon^2$, no state can have fidelity as large as $1-(\epsilon/2)^2$ with
more than one member of ${\cal M}_{\rm max}$ \footnote{If
$\ket{\beta}$ had fidelity as large as $1-(\epsilon/2)^2$ with both
$\ket{\varphi_k}$ and $\ket{\varphi_l}$, by the chain rule for fidelities, the
fidelity between $\ket{\varphi_k}$ and $\ket{\varphi_l}$ would have to be at
least $1-\epsilon^2$.}. Thus the regions of states with fidelity
at least $1-(\epsilon/2)^2$ with different elements of ${\cal M}_{\rm max}$ can
not intersect. One may therefore determine an upper limit on the number of
states in ${\cal M}_{\rm max}$ by dividing the volume of the region of
normalized states by the volume of the region of states that has fidelity at
least $1-(\epsilon/2)^2$ with some state $\ket{\varphi}$.

The region of allowed states is the surface of a hypersphere, and has volume
$2\pi^d/(d-1)!$. From Appendix \ref{volume}, the volume of a region with
fidelity at least $1-(\epsilon/2)^2$ is $2\pi^d (\epsilon/2)^{2d-2}/(d-1)!$.
Therefore the number of states in the set ${\cal M}_{\rm max}$ is no larger than
$(2/\epsilon)^{2d-2}$. In order to be able to perform the
entanglement transformations, we require fidelity at least $1-\epsilon^2$, where
$\epsilon=r\sqrt{d-1}$. Therefore, the communication required for this
non-constructive coding scheme is no more than
\begin{equation}
\label{third}
(2d-2)\log(2/r\sqrt{d-1}).
\end{equation}

We may place a lower bound on the communication required in a similar way. To
do this, we divide the total volume of the region of normalized states by the
volume of the region of states with fidelity at least $1-\epsilon^2$ (with an
arbitrary state). Clearly, if the number of states in ${\cal M}$ were less than
this, then there would be at least some states that did not have fidelity at
least $1-\epsilon^2$ with any state in ${\cal M}$. The volume of normalized
states is $2\pi^d/(d-1)!$, whereas the region of states with fidelity at least
$1-\epsilon^2$ has volume $2\pi^d \epsilon^{2d-2}/(d-1)!$. Thus the total
number of states can be no less than $(1/\epsilon)^{2d-2}$.

Taking $\epsilon=r\sqrt{d-1}$, the classical communication can be no less than
\begin{equation}
\label{fourth}
(2d-2)\log(1/r\sqrt{d-1}).
\end{equation}
The communication required for the non-constructive method is close to this, as
it is no more than $2d-1$ bits larger. In addition, the lower bound
\eqref{fourth} is similar to the first term in Eq.\ \eqref{second}; therefore,
provided the first term in \eqref{second} is dominant, the explicit method
that we described earlier is close to optimal.

As the classical communication for the rest of the scheme is $\log[d^2(d-1)]$,
for exact remote state preparation schemes of this type, the total
communication used can not be less than
\begin{equation}
\label{lower}
\log[d^2(d-1)]+(2d-2)\log(1/r\sqrt{d-1}),
\end{equation}
and there will be a scheme that uses communication of
\begin{equation}
\label{upper}
\log[d^2(d-1)]+(2d-2)\log(2/r\sqrt{d-1}).
\end{equation}
These expressions are plotted for the case of $d=2$ in Fig.\ \ref{fig}. There is
only a few bits difference between \eqref{upper} and \eqref{lower}, and the
explicit scheme given before requires communication that is greater than both
\eqref{upper} and \eqref{lower}.

It must be emphasised that the lower bound \eqref{lower} is not a lower bound
for arbitrary exact remote preparation schemes. One reason is that it was
derived from the requirement that a state must be specified with fidelity
$1-r^2(d-1)$. In order for it to be possible to apply the entanglement
transformation from $\ket{\Alpha}$ to $\ket{\Psi}$, it is only necessary that
$\vec\alpha^2\prec\vec\psi^2$. The volume of states satisfying this
condition will, in most cases, be larger, so it will be possible to specify the
state with less communication (though more communication will be required for
the state transformation). It is also possible that there may be some very
different remote state preparation scheme that uses less communication.

\section{Schmidt Number Required}
\label{sec:schmidt}
It is possible to obtain stronger results for the Schmidt number of the
entangled state. For the RSP scheme outlined above the Schmidt number of the
entangled state used must be maximal. It is possible to prove that this is
necessary for arbitrary exact RSP schemes as follows. First, note that the
above exact RSP scheme is equivalent to a local measurement performed in
subsystem $A$, followed by a unitary transformation applied in subsystem $B$
that is based on information communicated from subsystem $A$.

This is not the most arbitrary RSP scheme possible. In general, one may add
local ancillas, perform local unitary transformations, local general
measurements, and two-way communication. The POVMs used in each subspace may
depend on the results of previous measurements. Let the initial state be
\begin{equation}
\label{smaller}
\ket{\Alpha} = \sum_{k=0}^{d'-1} \alpha_k \ket{k}\ket{k},
\end{equation}
where $d'<d$. Because the local unitary transformations and measurement
operators on subsystem $A$ commute with those on subsystem $B$, we may combine
the operators on subsystem $A$ into a single operator $M_A(\beta,\vec\phi)$.
This operator may depend on the state to be prepared, $\ket{\beta}$, as well as
the results of measurements, $\vec\phi$. The vector $\vec\phi$ contains the
results of measurements performed in both subsystems. We allow $\vec\phi$ to
contain real numbers resulting from measurements in both subsystems (even though
these results can not be communicated with finite classical communication), as
this does not make the RSP scheme less general. We also combine the operators on
subsystem $B$ into a single operator $M_B(\vec n,\vec\phi)$. This operator also
may depend on the results of measurements $\vec\phi$, as well as additional
information $\vec n$ communicated from subsystem $A$.

After performing operation $M_A(\beta,\vec\phi)$, the reduced density matrix in
subsystem $B$ is
\begin{equation}
\rho \otimes \rho_{\rm anc},
\end{equation}
where $\rho_{\rm anc}$ is the state of the ancilla for subsystem $B$. As the
ancilla for subsystem $B$ is initially unentangled, it can not be modified in
any way by $M_A(\beta,\vec\phi)$. In addition, although $\rho$ will depend on
$M_A(\beta,\vec\phi)$, it must still be orthogonal to $\ket{k}$ for $k>d'-1$.
Without loss of generality, we assume that it is possible to prepare any state
$\rho$, provided it is orthogonal to $\ket{k}$ for $k>d'-1$.
In order to obtain perfect RSP, we require
\begin{equation}
\label{perfect}
\ket{\beta}\bra{\beta} = {\rm Tr}_{\rm anc} \left[ M_B(\vec n,\vec\phi)
(\rho \otimes \rho_{\rm anc})M\dg_B(\vec n,\vec\phi) \right].
\end{equation}
If Eq.\ \eqref{perfect} holds for $\rho$ and $\rho_{\rm anc}$, there must be
pure states for which it holds. Therefore we may take these states to be
$\ket{\psi}$ and $\ket{\psi_{\rm anc}}$. Eq.\ \eqref{perfect} then becomes
\begin{equation}
\ket{\beta}\otimes\ket{\psi'_{\rm anc}} = M_B(\vec n,\vec\phi)
\ket{\psi}\otimes\ket{\psi_{\rm anc}},
\end{equation}
where $\ket{\psi'_{\rm anc}}$ is the final state of the ancilla.

In order to obtain $\ket{\beta}$, for any given measurement results $\vec\phi$,
one may adjust $\ket{\psi}$ and the communicated information $\vec n$.
An arbitrary pure $d$-dimensional state $\ket{\beta}$ is equivalent to a point
on a $2d-1$ dimensional hypersphere (one dimension may be omitted because we
may take $\beta_0$ to be real). Because $\ket{\psi_{\rm anc}}$ is fixed, and
$\ket{\psi}$ is orthogonal to $\ket{k}$ for $k>d'-1$, the state $\ket{\psi}
\otimes\ket{\psi_{\rm anc}}$ is equivalent to a point on a $2d'-1$ dimensional
hypersphere. Since there is only a finite set of messages that may be
communicated $\vec n$, the set of states obtained by varying $\vec n$ and
$\ket{\psi}$ can only correspond to a $2d'-2$ dimensional space, and cannot fill
the $2d-2$ dimensional space corresponding to the set of states $\ket{\beta}$.

Therefore, even if it is possible to prepare an arbitrary $d'$-dimensional
state and perform one of a finite number of operations, it is not possible to
prepare an arbitrary $d$-dimensional state. Thus it is not possible to exactly
prepare an arbitrary $d$-dimensional state if the resource state has lower
Schmidt number.

\section{Conclusions}
\label{sec:conc}
We have given an explicit scheme for performing exact RSP using an arbitrary
entangled state with maximal Schmidt number and classical communication that
is close to optimal for schemes of this type. The scheme is a three step
process, involving an entanglement transformation, followed by a disentangling
measurement and a final unitary operation to obtain the exact state.

This method improves on given in Ref.\ \cite{ye} in two main ways: \\
1. The communication required for the entanglement transformation is less than
$2\log d$, as compared to $\log d!$ for Ref.\ \cite{ye}. \\
2. We have given an explicit method for determining the final unitary
operation. \\

The majority of the communication is required for the final unitary operation.
The communication required for this step is slightly superlinear in $d$, whereas
the communication required for the first two steps is logarithmic in $d$. This
communication is close to optimal, provided the remote state preparation scheme
is of this type; however, we have not eliminated the possibility that some more
general remote state preparation scheme may require less communication.

This remote state preparation scheme also requires that the Schmidt number of
the initial entangled state be maximal. We have proven that this is necessary
even for an arbitrary remote state preparation scheme.

\appendix
\section{Distance and fidelity}
\label{dist}
Consider two states that satisfy
\begin{equation}
\label{appdist}
\big\| \ket{\beta}-\ket{\tilde\beta'}\big\|\le \epsilon,
\end{equation}
where $\ket{\tilde\beta'}$ is not necessarily normalized. The state
$\ket{\tilde\beta'}$ may be expressed as $\ket{\tilde\beta'}=
a\ket{\beta}+b\ket{\beta^\perp}$, where $\ket{\beta^\perp}$ is orthogonal to
$\ket{\beta}$. Then Eq.\ \eqref{appdist} is equivalent to
$|1-a|^2+|b|^2\le \epsilon^2$, which implies
\begin{equation}
|a|\ge 1-\sqrt{\epsilon^2-|b|^2},
\end{equation}
and
\begin{equation}
\frac{|b|}{|a|} \le \frac{|b|}{1-\sqrt{\epsilon^2-|b|^2}}.
\end{equation}
The right-hand side of this expression is minimized for
$|b|^2=\epsilon^2-\epsilon^4$, giving
\begin{equation}
\frac{|b|^2}{|a|^2} \le \frac{\epsilon^2}{1-\epsilon^2}.
\end{equation}
In turn this implies
\begin{equation}
\frac{|a|^2}{|a|^2+|b|^2} \ge 1-\epsilon^2.
\end{equation}
If $\ket{\beta'}$ is the normalized state corresponding to $\ket{\tilde\beta'}$,
then
\begin{equation}
\big| \braket{\beta}{\beta'}\big|^2 = \frac{|a|^2}{|a|^2+|b|^2}
\end{equation}
Therefore $\| \ket{\beta}-\ket{\tilde\beta'}\|\le \epsilon$ implies
that $|\braket{\beta}{\beta'}|^2\ge 1-\epsilon^2$.

\section{Majorization and fidelity}
\label{major}
In this appendix it is shown that $\vec\alpha^2\prec\vec\psi^2$ is satisfied if
$\psi_0^2\ge 1-(d-1)r^2$. The majorization condition
$\vec\alpha^2\prec\vec\psi^2$ is equivalent to
\begin{equation}
\sum_{k=0}^p {^\downarrow}\psi_k^2 \ge \sum_{k=0}^p {^\downarrow}\alpha_k^2,
\end{equation}
where the down arrow indicates that the coefficients are sorted into descending
order. To show this result, note that, because the ${^\downarrow}\psi_k^2$ are
in descending order,
\begin{equation}
\frac 1p\sum_{k=1}^p {^\downarrow}\psi_k^2 \ge \frac 1{d-p-1}\sum_{k=p+1}^{d-1}
{^\downarrow}\psi_k^2.
\end{equation}
Multiplying on both sides by $d-p-1$ and adding $\sum_{k=1}^p
{^\downarrow}\psi_k^2$ gives
\begin{equation}
\frac {d-1}p \sum_{k=1}^p {^\downarrow}\psi_k^2 \ge (1-{^\downarrow}\psi_0^2).
\end{equation}
In turn this gives
\begin{equation}
\sum_{k=0}^p {^\downarrow}\psi_k^2 \ge {^\downarrow}\psi_0^2
\frac {d-p-1}{d-1}+\frac p{d-1}.
\end{equation}

The substituting the inequality $\psi_0^2\ge 1-(d-1)r^2$ (and using
${^\downarrow}\psi_0^2\ge \psi_0^2$) gives
\begin{equation}
\sum_{k=0}^p {^\downarrow}\psi_k^2 \ge 1-(d-p-1)r^2.
\end{equation}
Because $\alpha_k^2\ge r^2$, it is also the case that
\begin{equation}
1-(d-p-1)r^2 \ge \sum_{k=0}^p {^\downarrow}\alpha_k^2,
\end{equation}
thus giving
\begin{equation}
\sum_{k=0}^p {^\downarrow}\psi_k^2 \ge \sum_{k=0}^p {^\downarrow}\alpha_k^2.
\end{equation}
Hence the inequality $\psi_0^2\ge 1-(d-1)r^2$ is sufficient to imply the
majorization relation $\vec\alpha^2\prec\vec\psi^2$.

\section{Efficient coding}
\label{efficient}
If the numbers $n_k^r$ and $n_k^c$ record the absolute values of the real and
imaginary parts of $\beta_k$, and the interval $[0,1]$ is divided into $D'$
subintervals, then the fidelity is
\begin{equation}
\big| \braket{\beta}{\beta'}\big|^2\ge 1-\frac{2d-1}{4D'^2}.
\end{equation}
The number of subintervals should therefore be taken to be
\begin{equation}
D' = \left\lceil \sqrt {\frac {2d-1}{4r^2(d-1)} } \right\rceil .
\end{equation}
The number of bits required to encode the length of the bit-strings
representing each of the numbers $n_k^r$ and $n_k^c$ is
$\log \lceil \log D' \rceil$. In addition $\beta_0$ is taken to be real, and
we require $2d-1$ bits to record the signs of the real and imaginary parts of
$\beta_k$. The total communication is therefore
\begin{align}
& (2d-1)\log \lceil \log D' \rceil + \lceil \log (n_0^r-1) \rceil \nn \\
& \quad + \sum_{k=1}^{d-1} \big[ \lceil \log (n_k^r-1) \rceil +
\lceil \log (n_k^c-1) \rceil \big] +(2d-1) \nn \\
& \le (2d-1)\log \lceil \log D' \rceil + \lceil \log (D\beta_0) \rceil \nn \\
& \quad +  \sum_{k=1}^{d-1} \big[ \lceil \log (D{\rm Re}\beta_k) \rceil +
\lceil \log (D{\rm Im}\beta_k) \rceil \big] +(2d-1) \nn \\
& \le (2d-1)\log \lceil \log D' \rceil + (2d-1)\left(2+\log\frac {D'}
{\sqrt{2d-1}} \right)\nn \\
& \le (2d-1)\left[ -\log (r\sqrt{d-1}) + \log \lceil \log D' \rceil + 2\right].
\end{align}

\section{Volume of region of states}
\label{volume}
Here we consider the problem of determining the volume of the region of states
$\ket{\beta}$ for a given $\ket{\varphi}$ that satisfy
$|\braket{\varphi}{\beta}|^2\ge 1-\epsilon^2$. To do this, we write the state
$\ket{\beta}$ in the form
\begin{equation}
\ket{\beta}=e^{i\phi} \cos\theta \ket{\varphi} + \sin\theta \ket{\varphi^\perp},
\end{equation}
where $\ket{\varphi^\perp}$ is a state perpendicular to $\ket{\varphi}$. Every
state may be represented in this way when the ranges of $\phi$ and $\theta$ are
$[-\pi,\pi]$ and $[0,\pi/2]$, respectively. The condition
$|\braket{\varphi}{\beta}|^2\ge 1-\epsilon^2$
implies that $|\sin\theta|\le \epsilon$. The volume of states is given by
\begin{equation}
V=\int_0^{\arcsin\epsilon} d\theta\int_{-\pi}^{\pi} d\phi\cos\theta 
S_{2d-2}(\sin\theta),
\end{equation}
where 
\begin{equation}
\label{hyper}
S_n(r)=\frac{2\pi^{n/2}}{\Gamma(n/2)}r^{n-1}
\end{equation}
is the surface area of a hypersphere. Integrating over $\phi$ and using
\eqref{hyper} gives
\begin{align}
V&=\frac{4\pi^d}{(d-2)!}\int_0^{\arcsin\epsilon}
\cos\theta\sin^{2d-3}\theta d\theta \nn \\
&=\frac{4\pi^d}{(d-2)!} \left[ \frac{\sin^{2d-2}\theta}{2d-2}
\right]_0^{\arcsin\epsilon} \nn \\
&=\frac{2\pi^d}{(d-1)!} \epsilon^{2d-2}.
\end{align}
Note that using $\epsilon=1$ recovers the formula for
the surface area of a $2d$ dimensional hypersphere, which gives the total volume
of normalized states.

\acknowledgments
I am grateful for valuable discussions with Patrick Hayden, Barry C. Sanders
and Guifr\'e Vidal. This research has been supported by Alberta's informatics
Circle of Research Excellence (iCORE).

\end{document}